# Sub-ns intrinsic response time of PbS nanocrystal IR-photodetectors

*Andre Maier[1,2,#], Fabian Strauß[1,2], Pia Kohlschreiber[1,2], Christine Schedel[1], Kai Braun[1,2,#], Marcus Scheele[1,2,#]*

1. Institute of Physical and Theoretical Chemistry, Universität Tübingen, Auf der Morgenstelle 18, D-72076 Tübingen, Germany

2. Center for Light-Matter Interaction, Sensors & Analytics LISA+, Universität Tübingen, Auf der Morgenstelle 15, D-72076 Tübingen, Germany

\# Corresponding authors: andre.maier@uni-tuebingen.de, kai.braun@uni-tuebingen.de, marcus.scheele@uni-tuebingen.de

## ABSTRACT

Colloidal nanocrystals (NCs), especially lead sulfide NCs, are promising candidates for solution-processed next-generation photodetectors with high-speed operation frequencies. However, the intrinsic response time of PbS-NC photodetectors, which is the material-specific physical limit, is still elusive, as the reported response times are typically limited by the device geometry.

Here, we use the two-pulse coincidence photoresponse technique to identify the intrinsic response time of 1,2-ethanedithiol-functionalized PbS-NC photodetectors after fs-pulsed 1560 nm excitation. We obtain an intrinsic response time of ~1 ns, indicating an intrinsic bandwidth of ~0.55 GHz as the material-specific limit. Examination of the dependence on laser power, gating, bias, temperature, channel length and environmental conditions suggest that Auger recombination, assisted by NC-surface defects, is the dominant mechanism. Accordingly, the intrinsic response time might further be tuned by specifically controlling the ligand coverage and trap states. Thus,

PbS-NC photodetectors are feasible for gigahertz optical communication in the third telecommunication window.

# INTRODUCTION

The conversion of light into electrical signals is the basis of numerous technological applications, ranging from high-speed optical communications, video imaging and night-vision to gas sensing and biomedical imaging.[1–3]

Especially the detection of infrared (IR) light is of superior interest for next-generation optoelectronic devices, being beyond the wavelength range of Si photodetectors.[4] Typical IR-photodetectors consist of costly, epitaxially grown narrow bandgap semiconductors, such as InGaAs and HgCdTe, rendering IR-photodetectors significantly more expensive than Si sensors.[4]

Highly promising candidates for IR-photodetector materials are novel 2D materials such as graphene or black phosphorous.[1,5–8] However, these 2D materials feature some inherent drawbacks, e.g., low absorption, laborious fabrication and high dark current.[4,9]

Another promising class of materials are colloidal NCs, facilitating solution-processable low-cost photodetectors with large photoabsorption cross-sections and a compatibility with flexible substrates.[10–14] Especially lead sulfide (PbS) NCs are high-potential candidates due to their size-tunable absorption throughout the near-IR region of ∼600–3000 nm.[10,15–18]

Significant progress has been achieved in recent years regarding the PbS-NC photodetector's key figures of merit – the detectivity and response time.[2,3,19,20] Recently, high-speed PbS-NC photodetectors with promising response times down to 7–10 ns have been presented,[19,21] while a huge range of response times up to seconds have been previously reported.[15,22,23]

As those response times were determined by transient photocurrent measurements, they represent extrinsic response times. The extrinsic response time combines the intrinsic material properties with limitations due to parasitic capacitances, device geometry and resistance-capacitance (*RC*) time.[6,24–27] Hence, the intrinsic response time of PbS-NC photodetectors, i.e., their material-specific physical limit, is in fact still unknown.

Here, we apply two-pulse coincidence (2PC) photoresponse measurements via asynchronous optical sampling (ASOPS) to reveal the intrinsic response time $\tau$ of semiconductor-metal IR-photodetectors composed of colloidal PbS-NCs surface-functionalized with 1,2-ethanedithiol (EDT). This time-resolved pump-probe method is ideally suited to determine the intrinsic response time of photodetectors, which cannot be measured by direct electronic measurements.[24,28] It is also known as photoresponse autocorrelation technique[24,27,29] or two-pulse photoresponse correlation[30–32].

The photodetector channel is excited with two identical laser pulses separated by a delay time $\Delta t$. The corresponding photoresponse is recorded as a function of $\Delta t$ between the pulses. The photoresponse is reduced at zero delay $\Delta t=0$ when both pulses arrive together, due to a non-linear power dependence of the photoresponse. For larger $\Delta t$, the photoresponse exponentially recovers, as the photoexcited charge carriers in the photodetector relax to the ground state and can be excited again. The resulting symmetric 2PC dip around $\Delta t$ indicates the time the photodetector channel requires to return to its equilibrium state, which is the intrinsic response time $\tau$.[27]

Usually, the delay $\Delta t$ between the two laser pulses is generated by splitting the beam of one laser into two optical paths, where one is gradually varied by introducing a mechanical delay line, before recombining the two laser paths.[6,8,24,27,31–37] Thus, typical delay windows cover a range of only a few ps to ±250 ps[9,32–35,37], sometimes up to the ns timescale[5,8]. Furthermore, the 2PC measurements are acquired point-by-point by gradually varying $\Delta t$ and recording the corresponding photoresponse, resulting in ~hours of acquisition time for a single 2PC measurement.

Instead of a mechanical delay stage, we use two electronically coupled femtosecond (~65 fs) fiber lasers ($\lambda_{exc}$ = 1560 nm) with a slight offset in repetition rates ($f_{rep}$ = 100 MHz and $f_{rep} \pm \Delta f$), generating pump and probe pulses with an accumulating delay time $\Delta t$, sweeping over the entire period ($1/f_{rep}$), known as asynchronous optical sampling (ASOPS).[38–40]

2PC measurements in combination with ASOPS have previously been reported, enabling delay windows of ±140 ps to ±5 ns.[7,28,41,42] However, optically chopping of the laser pulses was required to enable lock-in detection. This can be circumvented using a lock-in amplifier with a periodic waveform analyzer (PWA) function, locked at the detuning frequency $\Delta f$. This results in a delay window of 10 ns with a scan resolution of 10 fs within only 10 ms of experimental acquisition time, as the periodic 2PC curves can be recorded in real time as a function of $\Delta t$ and averaged over seconds to obtain a high signal-to-noise ratio. Details can be found in the Methods and the schematic illustrations in Figure S1,S2.

We unravel the intrinsic response time of the EDT-functionalized PbS-NC photodetectors to be $\tau \sim 1$ ns, which is not transit- but recombination-limited. The dominant mechanism of the photocarrier recombination is identified to be trap-assisted Auger recombination, strongly depending on the density of trap states. The determined intrinsic response times of $\tau \sim 1$ ns indicates that gigahertz electrical switching (~0.55 GHz bandwidth) is feasible.

# RESULTS AND DISCUSSION

We used EDT-exchanged PbS-NCs with a diameter of 6.0±0.7 nm,[43] exhibiting the first absorption peak at around 1560 nm, to fabricate colloidal NC IR-photodetectors from solution (**Figure 1a,b** and Figure S3,S4). **Figure 1c** shows typical *I-V* characteristics of an EDT-treated PbS-NC photodetector in the dark and under quasi-CW laser illumination ($\lambda_{exc}$ = 1560 nm). The linear *I-V* dependence indicates an Ohmic behavior of the detectors. Under excitation, the current *I* increases by a factor of >10$^3$. The conductivity $\sigma$ of the PbS-NC channels are calculated as $\sigma_{dark}$ ~ $6 \times 10^{-6}$ S/m and $\sigma_{photo}$ ~ $7 \times 10^{-3}$ S/m for the dark- and photoconductivity, respectively. The quasi-CW laser illumination is performed by illumination with a pulsed laser with ~65 fs pulses, $f_{rep}$ = 100 MHz, $P_{average}$ ~ 180 mW and sampling of the corresponding photocurrent on the timescale of seconds. In contrast, no conductivity and photoresponse were observed for the native oleic acid-functionalized PbS-NC (PbS-OA) photodetectors (Figure S5).

**Figure 1d** displays the photocurrent as a function of the average laser power under quasi-CW illumination. We observe a sublinear dependence of the photocurrent on the laser power, which can be fitted by a power law scaling with the photoresponse ~$P^{0.4\pm0.09}$, agreeing with other materials exhibiting a sublinear behavior.[8,27,30,31,33] This nonlinearity is a result of the saturable absorption of PbS-NCs[44] and it is a prerequisite for the 2PC photoresponse technique, which relies on a nonlinear power dependence of the photoresponse.[24]

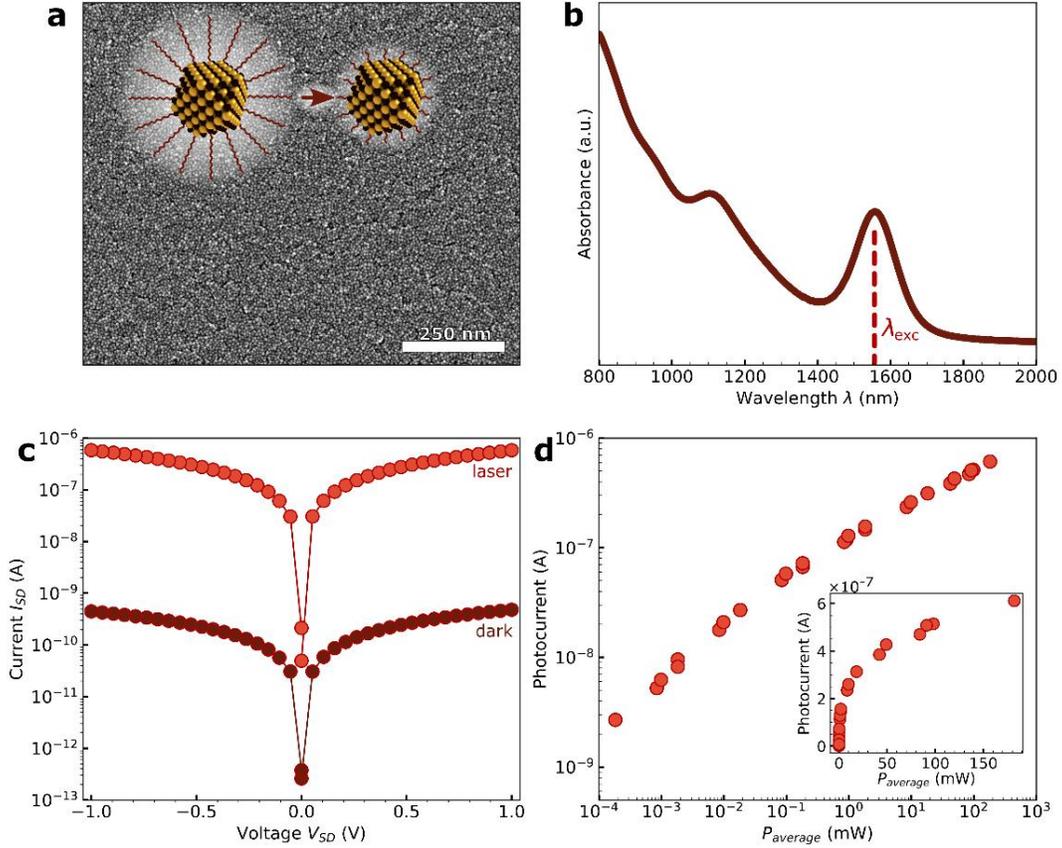

**Figure 1: Colloidal PbS-NC Photodetectors.** (**a**) Scanning-electron micrograph of the colloidal 6 nm PbS-NCs functionalized with EDT. The inset schematically illustrates the ligand exchange process of the native oleic acid ligands with EDT. High-resolution micrographs are given in Figure S4. (**b**) Absorption spectrum of the native PbS-NCs, showing a first excitonic transition at 1557 nm, matching the excitation wavelength of $\lambda_{exc}$ = 1560 nm. (**c**) Typical *I-V* characteristics of the colloidal PbS-EDT photodetector in the dark and under pulsed laser illumination. The photodetector is based on a PbS-EDT thin film ($h$ = 48±3 nm) on interdigitated electrodes with $L$ = 5.5 µm and $W$ = 1 cm on a Si/SiOx device (200 nm oxide). The current $I_{SD}$ increases by a factor of >$10^3$ under (quasi-CW) laser illumination (1560 nm, ~65 fs pulses, $f_{rep}$ = 100 MHz, $P_{average}$ ~ 180 mW). The connecting lines are guides to the eye. (**d**) Photocurrent as a function of the average laser power under quasi-CW illumination on a logarithmic scale. The inset shows the same data on a linear scale. The data can be fitted by a power law scaling with $P^{0.4\pm0.09}$. Channel biased with $V_{SD}$ = 1V, $V_G$ = 0 V.

**Figure 2a** illustrates the photoresponse measurement setup. The PbS-NC photodetector channels are illuminated by two laser pulses ($\lambda_{exc}$ = 1560 nm, ~65 fs pulse width, cross-polarized), whereas the delay time $\Delta t$ between the two pulses is varied by applying ASOPS. The main repetition frequency is $f_{rep}$ = 100 MHz, while one laser is slightly detuned by $\Delta f$ = 100 Hz. Consequently, the delay time $\Delta t$ gradually increases and is swept over the entire delay window of ±5 ns with a scan resolution of 10 fs within 10 ms acquisition time (temporal magnification factor of $10^6$ ($f_{rep}/\Delta f$))[45] without a mechanical delay line. The illuminated PbS-NC channels are connected to a lock-in amplifier, which is locked onto the detuning frequency $\Delta f$, to record the laser induced photovoltage as a function of the delay time $\Delta t$ between the two laser pulses. A constant source-drain bias $V_{SD}$ and gate voltage $V_G$ can be applied on the photodetector channels.

**Figure 2b** displays a typical 2PC measurement of an EDT-treated PbS-NC (PbS-EDT) photodetector, where the probe-induced change in photovoltage $\Delta V = V - V(\Delta t \gg 0)$ is plotted as a function of delay time $\Delta t$ (~ 90 mW per pulse). For $\Delta t \leq \tau$, the overall photoresponse is bleached due to the sublinear power dependence. Hence, the 2PC bleach directly reflects the ability of the detector to distinguish two temporally nearby pulsed excitations.[35] The time for the photocarrier generation can be neglected as it is generally very fast.[46]

We observe probe-induced photovoltage bleaches at $\Delta t = 0$ in the order of ~0.1–1 mV (see Supporting Section VI for details). By fitting the exponential decay of the bleach, the intrinsic response time $\tau$ can be determined, in this case $\tau$ ~ 1.0±0.1 ns (1/e value). For comparison, no 2PC bleach is obtained for a photodetector based on native PbS-OA NCs (blue curve).

In **Figure 2c**, the 2PC signal from Figure 2b is transformed to $\Delta V_{inv}$, defined as $|V - V(\Delta t \gg 0)|$, and logarithmically plotted as a function of $\Delta t$. The linearly decaying slopes for $|\Delta t| > 0$ indicate a mono-exponential recovery of the PbS-EDT channel on the time scale of $\tau$ = 1 ns. Using the approximation $f \approx 0.55/\tau$,[6,8,47] the observed intrinsic response time translates into an intrinsic (3dB) bandwidth of $f$ ~ 0.55 GHz, i.e., the maximum possible bandwidth if all external factors are negligible. See Supporting Section VII for a detailed discussion.

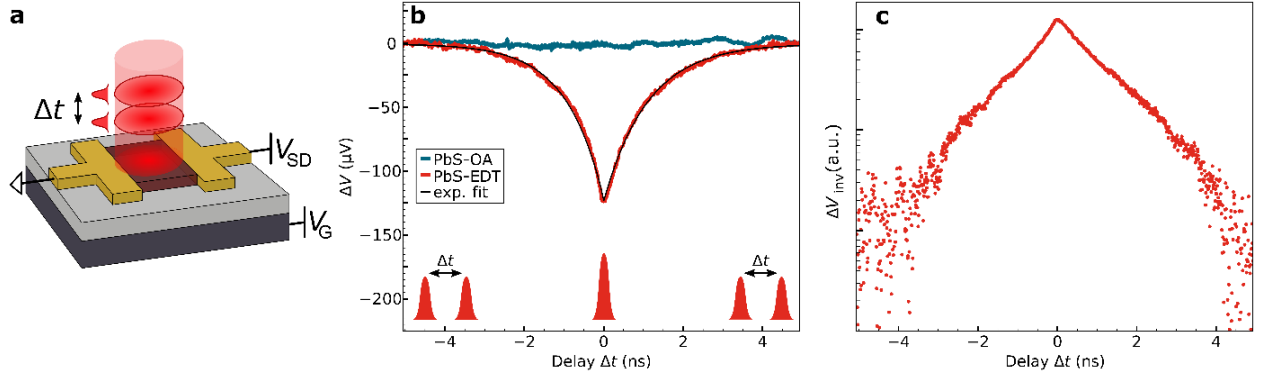

**Figure 2: Two-pulse coincidence (2PC) measurements.** (**a**) Schematic illustration of the 2PC setup. A PbS-NC channel is illuminated by two laser pulses ($\lambda_{exc}$ = 1560 nm, ~65 fs pulse width) with varying delay time $\Delta t$. The channel is biased with $V_{SD}$ and $V_G$. (**b**) 2PC measurements of a PbS-OA photodetector (blue) and a PbS-EDT photodetector (red), where the change in photovoltage $\Delta V$ is plotted as a function of $\Delta t$. For the PbS-EDT channel, near $\Delta t = 0$ a bleach is formed, exponentially recovering for large $\Delta t$. The exponential fit yields an intrinsic response time of $\tau \sim 1.0 \pm 0.1$ ns. No 2PC bleach is obtained for a photodetector based on PbS-OA NCs (blue curve). The measurements are normalized to the value for $\Delta t \gg 0$. (**c**) The data of the PbS-EDT channel from (b) transformed to $\Delta V_{inv} = |V - V(\Delta t \gg 0)|$, plotted as a function of $\Delta t$ on a logarithmic scale. The deviations at $\Delta t > \pm 4$ ns are caused by the baseline normalization. The channel dimensions are $L = 5.5$ µm, $W = 1$ cm, $h = 48 \pm 5$ nm, with $V_{SD} = 1$ V, $V_G = 0$ V, $P_{average} \sim 90$ mW per pulse, measured at room temperature under $N_2$ atmosphere.

We performed 2PC measurements of several channels at different pulse powers, biases, gate voltages, temperatures and channel lengths to understand the mechanism of the photoresponse of the PbS-EDT NC photodetectors.

A characteristic 2PC bleach is only observed for $V_{SD} > 0$ V. This is due to the absence of an intrinsic electric field in the investigated PbS-NC photodetectors, different to photodiodes with an intrinsic electric field.[13] We investigated the response time for different $V_{SD}$, which is roughly constant for all $V_{SD}$ (Figure S7a,b).

**Figure 3a** shows the measured 2PC signal $\Delta V$ as a function of the time delay $\Delta t$ of another device for different average laser powers $P_{average}$, ranging from ~$9 \times 10^{-4}$ mW to ~90 mW per pulse. As soon as a characteristic 2PC bleach is formed for high power, τ decreases from 1.8±0.02 ns to 1.45±0.09 ns by increasing $P_{average}$ from ~45 mW to ~90 mW. A similar behavior was also observed for another channel after air exposure (**Figure 3b**): τ corresponds to 1.80±0.8 ns, 1.29±0.29 ns and 0.83±0.02 ns for ~9 mW, ~45 mW and ~90 mW, respectively. As the signal-to-noise ratio at low power is not ideal, τ is subject to some uncertainty. τ speeds up with increasing $P_{average}$ and the photovoltage at large delay times $V(\Delta t >> 0)$ increases (**Figure 3c**). $V(\Delta t >> 0)$ values are extracted before normalization.

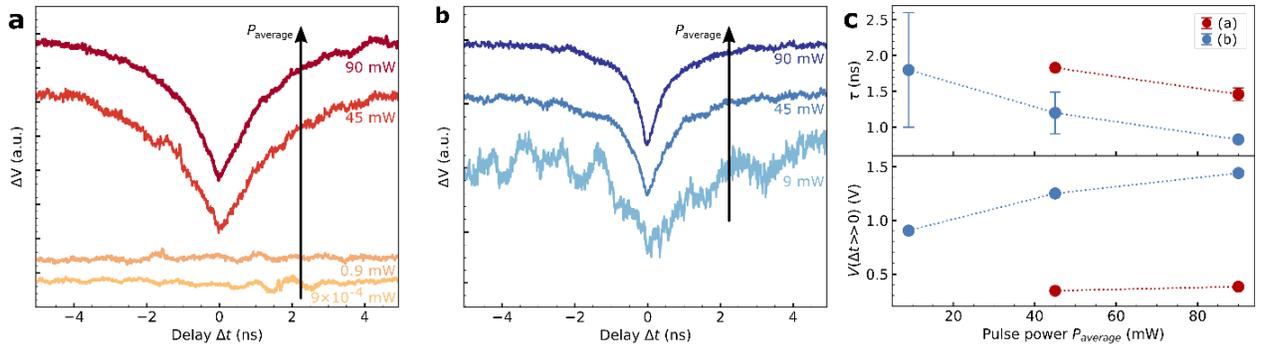

**Figure 3: Power-dependent 2PC measurements of PbS-NC channels.** (**a**) The 2PC signal $\Delta V$ as a function of $\Delta t$ of a PbS-EDT channel under $N_2$ atmosphere for different average laser powers $P_{average}$, ranging from ~9 × 10$^{-4}$ mW to ~90 mW per pulse. The response time τ slightly decreases from 1.8±0.02 ns to 1.45±0.09 ns by increasing $P_{average}$ from ~45 mW to ~90 mW. $V_{SD}$ = 1 V, $V_G$ = 0 V. (**b**) 2PC signal after air exposure for different $P_{average}$ of ~9 mW, ~45 mW and ~90 mW. The response times are 1.80±0.8 ns, 1.29±0.29 ns and 0.83±0.02 ns, respectively. $V_{SD}$ = 3 V, $V_G$ = 0 V. (**c**) The response time τ and photovoltage $V(\Delta t >> 0)$ as a function of $P_{average}$, corresponding to the data in (a) and (b). τ decreases with $P_{average}$, whereas $V(\Delta t >> 0)$ increases. The curves in (a,b) are normalized and offset for clarity, $V(\Delta t >> 0)$ values are extracted before normalization. Only data with 2PC bleach are shown. The error bars indicate the fitting uncertainty, and the connecting lines are guides to the eye. All measurements performed under $N_2$ at room temperature.

We observe the intrinsic response time $\tau$ to be insensitive to an applied gate voltage ($V_G$ = -10–10 V), temperature ($T$ = 10–300 K) and channel length ($L$ = 5.5–15 µm), as detailed in Figure S7 and Figure S8. The photovoltage $V(\Delta t \gg 0)$ linearly scales with $V_{SD}$, indicating the Ohmic behavior of the PbS-EDT channels. $V(\Delta t \gg 0)$ declines with decreasing $T$ and with increasing $L$ due to increasing resistance.

The PbS-EDT channels investigated under vacuum conditions show significantly longer response times of $\tau \sim$ 2 ns, whereas the same channels have shown $\tau \sim$ 1 ns response times under $N_2$ atmosphere. **Figure 4a** shows the evolving 2PC signal of a PbS-EDT channel after different treatments: directly after preparation under $N_2$ (~60 s in air), under vacuum (~4 × $10^{-5}$ mbar) and after air exposure (for 80 min) measured under $N_2$. The curves are normalized to overlap at $\Delta t$=0 and $\Delta t \gg 0$ for direct comparability. The initially sharp 2PC bleach with $\tau$ = 0.9±0.02 ns is broadened under vacuum conditions with $\tau$ = 2.0±0.2 ns and again sharpened with a further decreased response time of $\tau$ = 0.66±0.05 ns, as displayed in **Figure 4b**. The signal-to-noise ratio drastically increased under vacuum, whereas $V(\Delta t \gg 0)$ reversibly decreases (**Figure 4b**). The vacuum induced change of the PbS-EDT channel properties seems to be reversible, as the exposure to air drastically speeds up the response time. We investigated four different channels on the effect of air exposure on $\tau$ (Figure S10). The observed behavior was always similar: the intrinsic response time $\tau$ of the air-exposed channels was on average improved by a factor of 1.88, and $V(\Delta t \gg 0)$ increased by a factor of 2.1 (~doubled signal).

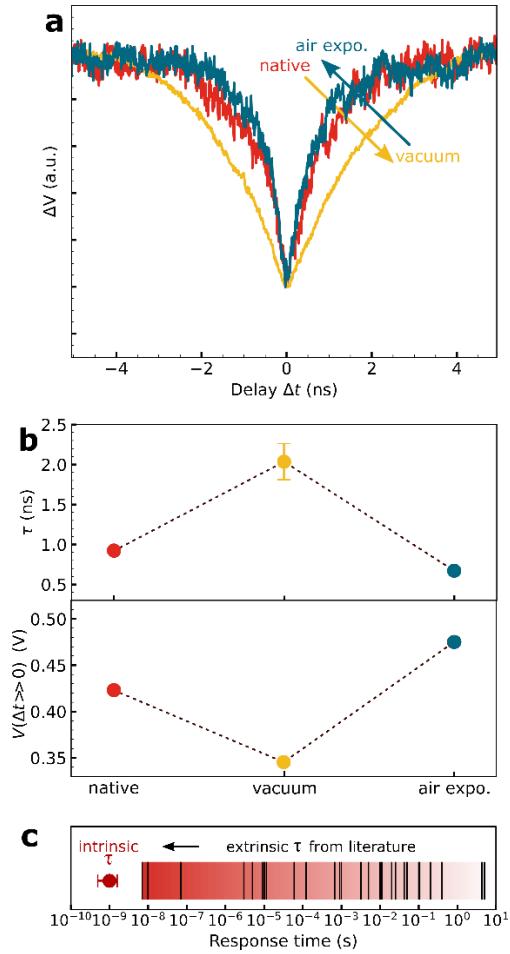

**Figure 4: Evolution of the 2PC signal after different treatments.** (**a**) 2PC curves of a PbS-EDT channel under $N_2$ (red), under vacuum (yellow) and after air exposure, measured under $N_2$ (blue). The curves are normalized to overlap at $\Delta t$=0 and $\Delta t$>>0 for direct comparison. (**b**) τ and $V(\Delta t$>>$0)$ of the data in (a). The connecting lines are guides to the eye. All measurements were performed at room temperature with $V_{SD} = 1$ V, $V_G = 0$ V and $P_{average}$ ~ 90 mW. (**c**) Comparison of typical response times of PbS-NC photodetectors reported in literature (black bars) along with the intrinsic response time of τ ~ 1 ns derived from this publication, indicating the physical limitation of the response speed of PbS-EDT photodetectors. Data taken from References [15,19,21,48].

Due to different dependencies of τ on various parameters, such as laser power, gating, bias, temperature and channel length, several mechanisms have been identified to be responsible for the ultrafast intrinsic response time of other photodetector devices: carrier recombination[6], hot carrier cooling[33] and supercollision[31,42] in graphene, Auger recombination in $MoS_2$[32], carrier transit in 2D oxyselenide crystals[35], recombination (transit) limitation for WSe$_2$ monolayers (multilayers)[27], recombination limitation in black phosphorous[7,8] and charge transfer within graphene/MoTe$_2$ heterostructures[9]. In the following, we discuss the mechanism responsible for the intrinsic ~1 ns response time of the reported PbS-NC photodetectors.

Photodetectors can return to their equilibrium state in different ways, since the photoexcited charge carriers have two possible decay paths. They can either drift out of the channel, including charge transfer to the metal electrodes (transit), resulting in a photocurrent/photovoltage, or they can recombine non-radiatively, introducing loss.[27] The total intrinsic decay rate $1/\tau$ denotes the photocarrier decay. It consists of the two components of the transit time and the recombination time, as follows [6]:

$$\frac{1}{\tau} = \frac{1}{\tau_{trans}} + \frac{1}{\tau_{recomb}} \quad . \tag{1}$$

The transit time $\tau_{trans}$ of a photocarrier can be calculated as $\tau_{trans} \sim L^2/\mu V_{SD}$. Thus, $\tau$ of a transit-limited channel is strongly bias- and channel length-dependent, which is both not valid for the reported PbS-EDT photodetectors (Figure S7a,b and S8). Using the estimated hole mobility derived via field-effect transistor measurements of $\mu(h^+) \sim 4 \times 10^{-5}$ cm$^2$/Vs (Supplementary Section IX), agreeing with typical values reported in literature[20,49], $\tau_{trans}$ can be estimated to be on the order of milliseconds, ruling out a transit-limited mechanism on the observed timescale. Consequently, a significant fraction of the photogenerated carriers recombines before leaving the PbS/metal interface region.[6,27]

As the observed $\tau$ is not transit-related, the recombination time on the ns-timescale seems to be the dominant mechanism in the PbS-EDT photodetectors, consistent with previous studies.[50–52] This is supported by the power-dependence of $\tau$. The inverse relationship between $\tau$ and pulse power (faster $\tau$ for higher power) points to trap-assisted Auger (TAA) recombination as the dominant recombination mechanism,[6,32,53,54] which is typical for PbS-NCs.[50,51,55] The trap-assisted recombination rate increases with the carrier density, which is in turn increased by higher excitation power.[2,55,56] Further, the negligible dependence of τ on $V_G$ hints towards a recombination-limited response time.[6] The temperature insensitivity of τ is also consistent with TAA recombination, observed for transition metal dichalcogenides (TMDCs).[32,53,57]

In most semiconductors, Auger processes play an important role for the carrier capture by defects at high carrier densities.[53] Due to the high pulse power, a high density of photocarriers can be assumed, which we estimate to $n \sim 10^{19}$ cm$^{-3}$ for the PbS-EDT films under photoexcitation (Supplementary Section XI).

However, another common mechanism of trap-assisted recombination in PbS-NC films with high trap density is the Shockley-Read-Hall (SRH) recombination, also observed to be invariant with temperature.[25,58,59] The recombination rates of individual mechanisms add up. In view of the monoexponential decay in the 2PC curves (**Figure 2c**), only one mechanism seems to be dominant. Whereas the SRH recombination rate depends on the trap density, TAA recombination depends on both the trap density and the excess carrier density.[60] Hence, due to the power-dependent – and, thus, carrier density-dependent – τ, the TAA recombination is considered the most probable mechanism responsible for the fast intrinsic response time of the PbS-EDT photodetectors.

Due to the large surface to volume ratio, trap states are intrinsically present in PbS-NCs, even in the absence of surface oxidation.[61–63] The trap states within the bandgap of PbS-NCs with a typical energy depth of 0.2–0.5 eV above the valence band originate from nonstoichiometric PbS facets or from an imbalance between excess PbS and coordinating ligands, i.e., Pb dangling bonds,[2,64] and can thus be manipulated through a specifically controlled ligand coverage.[51,61,65] Ligand exchange further increases the trap state density, commonly observed for EDT ligands[64], which facilitates carrier trapping and trap-assisted recombination.[66] In contrast, the surface trap passivation of CdSe and PbS-NCs resulted in an increased photocarrier lifetime, which was attributed to a reduced Auger recombination rate.[55,67]

As observed in **Figure 4a,b**, the vacuum treatment of the as-prepared PbS-EDT photodetectors reversibly decelerated the photocarrier recombination, which we attributed to a reduced trap-state density.[63] Exposing the PbS-EDT channels to air results in a partial oxidation of the surface of the not entirely covered PbS facets, typical for EDT-functionalized PbS-NCs.[43,64] This increases the trap state density which speeds up the trap-assisted photocarrier recombination.[12,43,54,64] Thus, the faster $\tau$ of the PbS-EDT photodetectors after air exposure strongly supports the hypothesis of a trap-assisted recombination mechanism. Such a trap-induced shortening of $\tau$ has also been observed for conventional GaAs photodetectors.[29]

**Figure 4c** displays the response times of typical PbS-NC photodetectors reported in literature,[15,19,21,48] corresponding to the extrinsic response time. For comparison, the intrinsic response time corridor of $\tau \sim 0.6\text{–}2$ ns, derived from this publication, is shown, indicating the physical limitation of the response speed of the PbS-EDT photodetectors (1560 nm excitation). Indeed, highly advanced PbS-NC photodetectors progressively approach the material-limited intrinsic response time.[19,21] However, it should be noted that the intrinsic response time might change with particle size, ligand coverage, etc.

The 2PC technique has been applied to reveal the intrinsic response time $\tau$ of a variety of materials. Various graphene devices have shown intrinsic response times in the range of $\sim 1.5\text{–}250$ ps[5,6,28,33,42], carbon-nanotubes of $\sim 0.6$ ps[30] and 2D oxyselenide crystals of $\sim 7$ ps[35]. Monolayer TMDC devices feature $\tau$ of $\sim 3\text{–}10$ ps, whereas thick multilayered TMDC devices exhibit slower $\tau$ of $\sim 1\text{–}10$ ns.[9,27,32] Similar nanosecond response times were observed for black phosphorous with $\tau \sim 1$ ns.[7] Thus, the reported colloidal PbS-EDT photodetectors feature an intrinsic response time comparable to highly sophisticated photodetector devices based on 2D materials as black phosphorous or WSe$_2$ and MoTe$_2$.[7,9,27] Moreover, the reported PbS-EDT photodetectors feature some significant benefits, such as their low-cost solution-processability, the tunable bandgap in the near-IR region and the high absorption,[2,18] along with the possibility of versatile chemical modifications by ligand exchange to further tune the response time of the photodetectors.

# CONCLUSION

We applied the two-pulse coincidence photoresponse technique, where the photoresponse is determined as a function of delay time $\Delta t$ between two exciting laser pulses, to reveal the intrinsic response time of EDT-treated PbS-NC photodetectors of $\tau \sim 1$ ns. This was possible due to the sublinear power-dependence of the detector's photoresponse, leading to a characteristic bleach of the 2PC curve at $\Delta t = 0$. By performing power-, bias-, gating- temperature- and channel length-dependent measurement, we identified trap-assisted Auger recombination as the limiting mechanism determining the intrinsic temporal response of the PbS-NC photodetectors. It was also shown that this material-specific parameter could be further enhanced by specifically controlling the ligand coverage and trap-state density.[51] The response time of $\tau \sim 1$ ns, down to even $\sim 660$ ps, is the fastest temporal response reported for PbS-NC photodetectors, indicating that the temporal responses reported in literature may still be device (e.g. *RC* time) limited.

## Supporting Information

Methods and experimental details, supporting sections including schematic explanations of two-pulse coincidence photoresponse measurements and ASOPS, characterization of the PbS nanocrystals and devices, probe-induced change in photovoltage, discussion of the bandwidth estimation, supporting 2PC measurements, field-effect transistor measurements and estimation of the carrier concentration.

## Acknowledgements

We acknowledge Claudius Riek from Zurich Instruments for fruitful discussions. We thank Elke Nadler for supporting SEM measurements. Financial support of this work has been provided by the European Research Council (ERC) under the European Union's Horizon 2020 research and innovation program (grant agreement No 802822) as well as the Deutsche Forschungsgemeinschaft (DFG) under grant SCHE1905/9-1.

## Author Contribution

A.M. performed the device fabrication, NC film preparation and characterization, developed the 2PC&ASOPS set-up, performed and analysed the measurements and wrote the manuscript. F.S. designed the device layout and supported the wafer fabrication as well as the ASOPS set-up development. P.K. performed the profilometry, optical microscopy and supported the thin film fabrication. C.S. supported the probestation set-up. A.M., K.B. and M.S. conceived the project. M.S. supervised the project. All authors have given approval to the final version of the manuscript.

## Competing interests

The authors declare no competing interests.

(16) Moreels, I.; Lambert, K.; Smeets, D.; de Muynck, D.; Nollet, T.; Martins, J. C.; Vanhaecke, F.; Vantomme, A.; Delerue, C.; Allan, G.; Hens, Z. Size-dependent optical properties of colloidal PbS quantum dots. *ACS Nano* **2009**, *3* (10), 3023–3030. DOI: 10.1021/nn900863a.

(17) Nikitskiy, I.; Goossens, S.; Kufer, D.; Lasanta, T.; Navickaite, G.; Koppens, F. H. L.; Konstantatos, G. Integrating an electrically active colloidal quantum dot photodiode with a graphene phototransistor. *Nat. Commun.* **2016**, *7*, 11954. DOI: 10.1038/ncomms11954.

(18) Tang, J.; Sargent, E. H. Infrared colloidal quantum dots for photovoltaics: Fundamentals and recent progress. *Adv. Mater.* **2011**, *23* (1), 12–29. DOI: 10.1002/adma.201001491.

(19) Vafaie, M.; Fan, J. Z.; Najarian, A. M.; Ouellette, O.; Sagar, L. K.; Bertens, K.; Sun, B.; García de Arquer, F. P.; Sargent, E. H. Colloidal quantum dot photodetectors with 10-ns response time and 80% quantum efficiency at 1,550 nm. *Matter* **2021**, *4* (3), 1042–1053. DOI: 10.1016/j.matt.2020.12.017.

(20) Xu, Q.; Meng, L.; Sinha, K.; Chowdhury, F. I.; Hu, J.; Wang, X. Ultrafast Colloidal Quantum Dot Infrared Photodiode. *ACS Photonics* **2020**, *7* (5), 1297–1303. DOI: 10.1021/acsphotonics.0c00363.

(21) Biondi, M.; Choi, M. J.; Wang, Z.; Wei, M.; Lee, S.; Choubisa, H.; Sagar, L. K.; Sun, B.; Baek, S. W.; Chen, B.; Todorović, P.; Najarian, A. M.; Rasouli, A. S.; Nam, D. H.; Vafaie, M.; Li, Y. C.; Bertens, K.; Hoogland, S.; Voznyy, O.; García de Arquer, F. P.; Sargent, S. H. Facet-Oriented Coupling Enables Fast and Sensitive Colloidal Quantum Dot Photodetectors. *Adv. Mater.* **2021**, *33* (33), 2101056. DOI: 10.1002/adma.202101056.

(22) Dong, C.; Liu, S.; Barange, N.; Lee, J.; Pardue, T.; Yi, X.; Yin, S.; So, F. Long-Wavelength Lead Sulfide Quantum Dots Sensing up to 2600 nm for Short-Wavelength Infrared Photodetectors. *ACS Appl. Mater. Interfaces* **2019**, *11* (47), 44451–44457. DOI: 10.1021/acsami.9b16539.

(67) Straus, D. B.; Goodwin, E. D.; Gaulding, E. A.; Muramoto, S.; Murray, C. B.; Kagan, C. R. Increased carrier mobility and lifetime in CdSe quantum dot thin films through surface trap passivation and doping. *J. Phys. Chem. Lett.* **2015**, *6* (22), 4605–4609. DOI: 10.1021/acs.jpclett.5b02251.

**TOC-FIGURE: For table of content only**

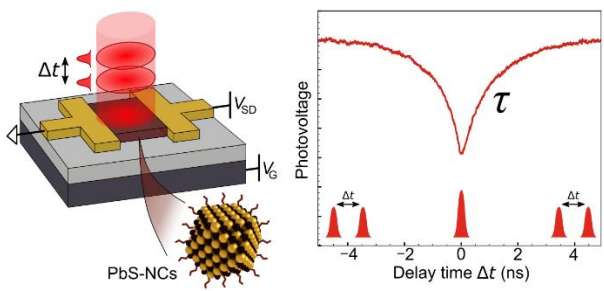

# SUPPORTING INFORMATION

## Sub-ns intrinsic response time of PbS nanocrystal IR-photodetectors

*Andre Maier[1,2,#], Fabian Strauß[1,2], Pia Kohlschreiber[1,2], Christine Schedel[1], Kai Braun[1,2,#], Marcus Scheele[1,2,#]*

3. Institute of Physical and Theoretical Chemistry, Universität Tübingen, Auf der Morgenstelle 18, D-72076 Tübingen, Germany

4. Center for Light-Matter Interaction, Sensors & Analytics LISA$^+$, Universität Tübingen, Auf der Morgenstelle 15, D-72076 Tübingen, Germany

\# Corresponding authors: andre.maier@uni-tuebingen.de, kai.braun@uni-tuebingen.de, marcus.scheele@uni-tuebingen.de

# METHODS

**Materials:**
All chemicals were used as received unless otherwise noted: Acetonitrile (99.9+%, extra dry, Acros Organics), 1,2-ethanedithiol (Sigma Aldrich), tetrachlorethylene (>99%, anhydrous, Acros Organics). Oleic acid (OA) stabilized PbS-NCs were synthesized according to Weidman et al.[1] Silicon/silicon dioxide (Si/SiO$_x$) wafer with 200 nm SiO$_x$ layer and n-doped Si were purchased from Siegert Wafer. Photoresist and developer (ma-N 405, ma-D 331/S, respectively) were purchased from Micro Resist Technology GmbH, Berlin.

**Device fabrication:**
For the electrode devices, a standard photolithography technique (negative tone resist) was used to pattern Au electrodes on Si/SiO$_x$ substrates (200 nm SiO$_x$). Au (~20 nm) and Ti (~5 nm) as an adhesion layer were thermally evaporated under high vacuum conditions (~8 × 10$^{-7}$ mbar). Ultrasonic-assisted lift-off in acetone removed the residual resist and metal layer. Interdigitated electrodes with gaps of 5.5–15 µm (channel length *L*) and a total channel width of *W* = 1 cm were realized.

The PbS-NC devices were coated as follows: 1) covering with 75 µL of a ~5 µM PbS-NC hexane solution and spin-coating after 30 s with a speed of 20 rps for 30 s, 2) covering with 150 µL of an EDT acetonitrile solution (5 mM) and spin-coating after 30 s at 20 rps for 30 s and 3) covering with 150 µL of acetonitrile and spin-coating after 30 s at 20 rps for 30 s. These three steps were repeated twice. The contact pads were cleaned of the NC film with a hexane-soaked cloth. Finally, the device is vacuum treated for 30 min. The PbS-OA devices were prepared by performing three times spin-coating step 1. All devices were prepared in a nitrogen filled glovebox.

**Nanocrystal and device characterization:**

Absorbance spectra of PbS-NCs in tetrachlorethylene were acquired with an UV-vis-NIR spectrometer (Cary 5000, Agilent Technologies). Scanning-electron microscopy of PbS-NC photodetector devices was performed with a HITACHI model SU8030 at 30 kV. The film thicknesses were characterized by profilometry (Dektak XT-A, Bruker). The investigated PbS-NC films exhibited thicknesses in the range of $h = 48\pm5$ nm – $55\pm3$ nm (roughness of 5–10%).

**Electrical measurements:**

All electrical measurements were conducted in a probe station (Lake Shore, CRX-6.5K). The contact pads were contacted with W-tips, connected to a source-meter-unit (Keithley, 2636 B). A back electrode worked as gate electrode. For two-point conductivity measurements, voltage sweeps in a certain range of ±1 V were applied and the current (as well as leak current) detected. The resistances of the illuminated PbS-NC photodetectors are measured to be $R = 1.6$–$2$ MΩ. For FET measurements (bottom-gate, bottom-contact configuration), a constant source-drain voltage $V_{SD}$ was applied and $I_{SD}$ was measured, modulated by applied gate voltage $V_G$ sweeps. Using the gradual channel approximation, field effect mobilities $\mu$ were calculated (c.f. Equation S2).

**Opto-electrical characterization and 2PC measurements via ASOPS:**

All opto-electrical measurements were conducted in a probe station (Lake Shore, CRX-6.5K). For the 2PC measurements via ASOPS, we used the Optical Sampling Engine (OSE) from Menlo Systems GmbH, featuring two femtosecond Erbium fiber lasers with a wavelength of $\lambda = 1560$ nm, pulse width of ~65 fs and average power of ~90 mW per laser, synchronized with high-accuracy phase-locking electronics. The two pulse trains within one PM1560 (PANDA) fiber are cross-polarized to minimize interference at zero delay[2]. The laser spot was not focused in order to illuminate the entire photodetector channel area (~0.2 mm$^2$). The main repetition rate corresponds to $f_{rep} = 100$ MHz, whereas one laser is slightly detuned by $\Delta f = 100$ Hz. This results in a full delay window of $1/f_{rep} = 10$ ns with a scan resolution of ~10 fs within 10 ms of sweep time. Details are given in the Supplementary Sections S4 and S5. The detuning frequency $\Delta f = 100$ Hz was used as a trigger signal for a lock-in amplifier (UHFLI from Zurich Instruments). The photodetector channels were connected with 50 Ω matched W-tips and 40 GHz coaxial cables to the UHFLI lock-in amplifier with 1 MΩ input impedance. The 2PC photovoltage was measured using the in-built Periodic Waveform Analyzer (PWA) function of the UHFLI lock-in amplifier, locked at the detuning frequency $\Delta f$. Thus, the periodic signal of the ASOPS sweep with its accumulating delay can directly be recorded, allowing to measure the photovoltage as a function of $\Delta t$ in real time. The periodic 2PC curves were recorded with 1024 bins and averaged over up to several giga samples. The PWA operates as a high-speed digitizer synchronized to an oscillator and therefore captures every sample without any dead time while rejecting all non-periodic signal components. The channels were biased by applying a $V_{SD}$ or $V_G$ using the UHFLI (maximum range of ±10 V). During the transfer of the samples into the nitrogen-filled sample station, they were exposed to air for the shortest possible period of about 60 s. Measurements were performed under nitrogen atmosphere or under vacuum (~$4 \times 10^{-5}$ mbar). The temperature was controlled in the range of 10–300 K, using a Lake Shore temperature controller (model 336). After reaching 10 K, the measurements were repeated at 300 K, to verify the reversibility. The 2PC dip fitting was performed by exponentially fitting both sides of the dip using Equation (S1), where $A$ and $B$ are fitting parameters. The mean value ± standard deviation was calculated from the fits of both sides.

$$\Delta V = A \times e^{(-\Delta t/\tau)} + B \tag{S1}$$

For the quasi-CW steady-state photocurrent measurements, the channels were contacted to a source-meter-unit (Keithley, 2636 B) as described above and the channels were illuminated as described above using the 1560 nm fiber lasers of the OSE. The steady-state photocurrent was measured at an applied bias $V_{SD}$ over time with an acquisition time of 1 s. The average laser power was reduced by inserting fiber-based attenuation filters.

The average laser pulse power was reduced by inserting fiber attenuators (3 dB, 10 dB, 20 dB) in the fiber-based laser-setup.

# SUPPORTING SECTIONS

## Supporting Section I: Two-pulse coincidence photoresponse measurements

**Figure S1** schematically illustrates the principle of the 2PC technique. At overlapping laser pulses near $\Delta t = 0$, a dip is formed due to the non-linear power dependence, which exponentially recovers at large $\Delta t$. For $\Delta t$ larger than the intrinsic response time $\tau$, the channel is fully de-excited after the first pulse arrival before re-exciting by the second pulse. The two pulses generate two independent photoresponse signals creating the total photovoltage.

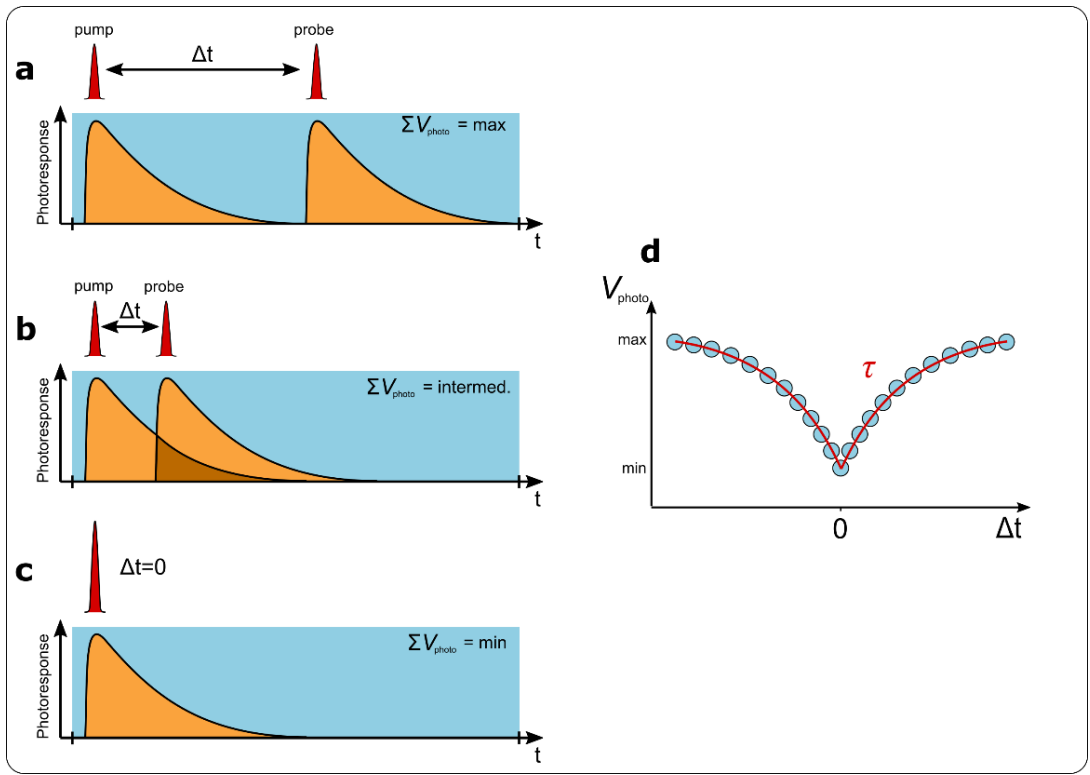

**Figure S1: Schematic illustration of the two-pulse coincidence photoresponse principle.** (**a-c**) The photoresponses as a function of time (orange curves) after laser excitation for an idealized system, where the pump pulse fully saturates the photoresponse. (**a**) For large delay times $\Delta t$, the pump and probe laser pulses generate two independent, identical photoresponses. Integration of the two signals (blue area) yields a maximum photoresponse $V_{photo}$ = max. (**b**) For intermediate $\Delta t$, the response is not fully vanished when the probe pulse arrives. Thus, the photoresponses overlap and integration yield a total photoresponse which is decreased by the overlapping region (dark orange area). $V_{photo}$ = intermediate. (**c**) For coinciding laser pulses at $\Delta t = 0$, only one photoresponse is created, as the increased power due to the probe pulse does not induce a higher response. The resulting total photoresponse is minimized, $V_{photo}$ = min. (**d**) Plotting the photoresponse $V_{photo}$ as a function of delay time results in a characteristic dip curve, due to the decreased photoresponse at $\Delta t = 0$, with an exponential recovery for large delay times. Exponential fitting (red curves) of the two recovery traces yields the intrinsic response time τ, which indicates the time scale the photodetector requires to return from its excited state to its equilibrium state.

## Supporting Section II: Asynchronous optical sampling (ASOPS)

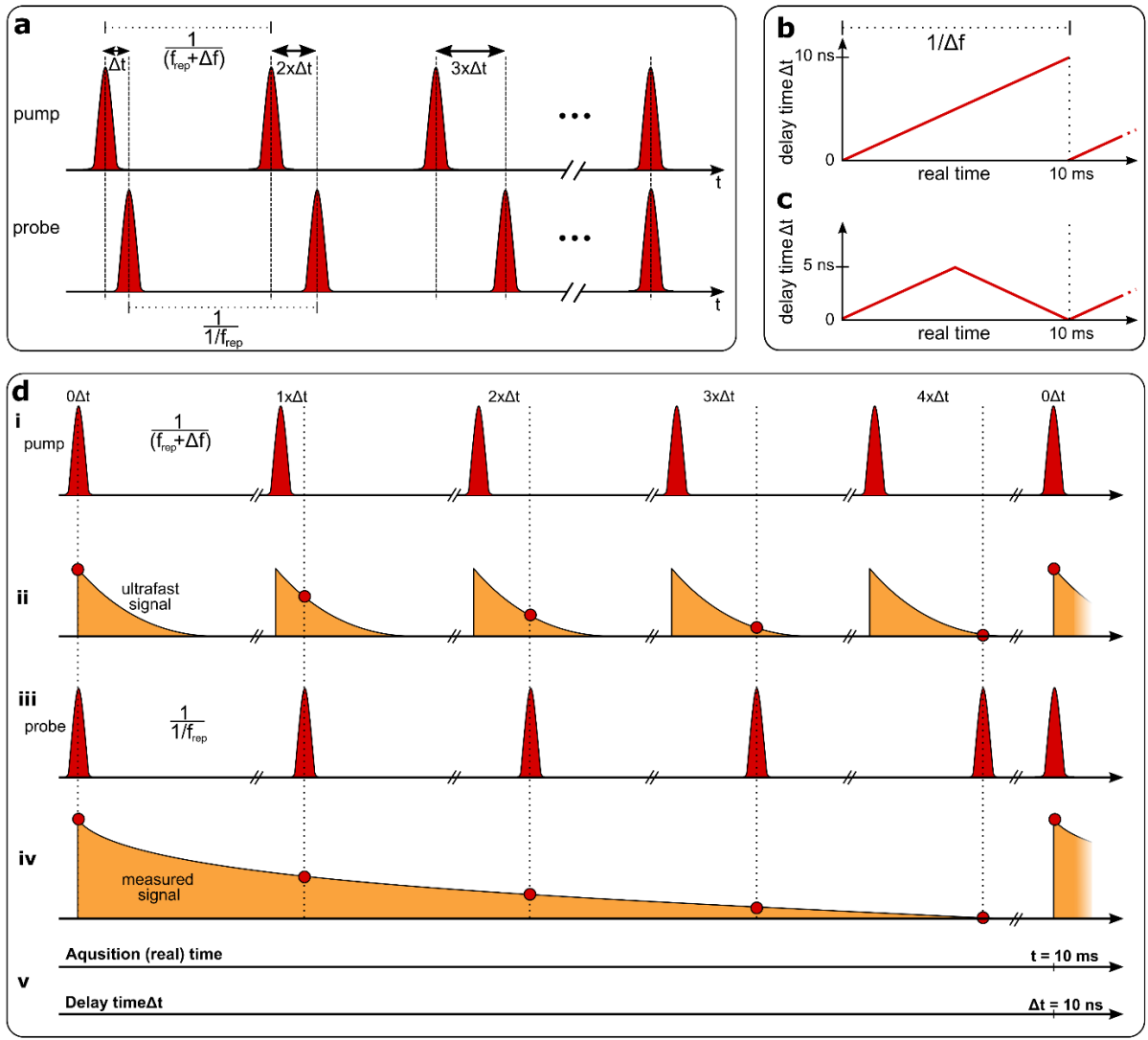

**Figure S2: Schematic illustration of the ASOPS working principle.** (**a**) Two lasers are electronically coupled and generate fs-laser pulses. The repetition rate $f_{rep}$ of the pump pulse is slightly enlarged by the detuning frequency $\Delta f$, whereas the probe pulses repeat with $f_{rep}$. Accordingly, the two pulses are relatively delayed to each other with increment $\Delta t$. The delay time $\Delta t$ gradually increases until both pulses again arrive simultaneously. (**b**) The full delay time ($\Delta t = 1/f_{rep}$) is scanned within the acquisition real time of $1/\Delta f$. For $f_{rep}$ = 100 MHz and $\Delta f$ = 100 Hz, the full delay time of $\Delta t$ = 10 ns is swept within $1/\Delta f$ = 10 ms, indicating a temporal magnification factor of $10^6$ ($f_{rep}/\Delta f$).[3] (**c**) In our experiment, both pump and probe laser pulses are identical and thus, interchangeable. Hence, the $\Delta t$ gradually increases to a maximum of 5 ns before it gradually decreases again, as pump and probe pulses are switched. For example, $\Delta t$ = 7 ns is identical to $\Delta t$ = -3 ns. (**d**) Time-transformation of ASOPS: (i) The pump pulses with detuned repetition rate ($f_{rep}+\Delta f$) excite the sample, inducing an ultrafast signal (orange curves in ii). (iii) The probe pulses with slightly lower repetition rate $f_{rep}$ probe the ultrafast signal after a certain delay time $\Delta t$, gradually increasing until both pump and probe lasers coincide again after the sweep time of $1/\Delta f$. Thus, the original ultrafast signal is measured in an iterative manner (iv), resulting in the time-transformation effect. (v) The full delay time of 10 ns can be time-transformed to an acquisition real time of 10 ms. Figure (c) adapted from References [4,5].

**Table S1: Characteristic parameters of the ASOPS principle.** The parameters are depicted for a detuning frequency of $\Delta f$ = 100 Hz. This results in full delay sweeps from 0 to 10 ns with 10 fs resolution within 10 ms of experimental acquisition time.

| Main repetition rate (of probe pulse) | $f_{rep}$ = 100 MHz |
|---|---|
| Delay time window | $1/f_{rep}$ = 10 ns |
| Repetition rate offset (detuning frequency) | $\Delta f$ = 100 Hz |
| Detuned repetition rate (of pump pulse) | $f_{rep,detun}$ = 100 MHz + $\Delta f$ |
| Scan resolution | $\Delta t_0 = \Delta f/(f_{rep} \times f_{rep}+\Delta f) \sim \Delta f/f_{rep}^2$ = 10 fs |
| Acquisition (real) time | $1/\Delta f$ = 10 ms |

## Supporting Section III: Characterization of the PbS-NCs

We used oleic acid (OA) capped PbS-NCs with a diameter of 6.0±0.7 nm to fabricate colloidal NC photodetectors from solution. A scanning electron micrograph and an optical absorption spectrum of the PbS-NCs are displayed in Figure 1a,b (main text). The NCs feature a first absorption peak at 1557 nm, corresponding to a band gap of $E_{gap}$ = 796 meV. A magnification of the first excitonic transition is given in **Figure S3a**. Using sizing-curves to the UV-Vis spectra,[6] the NC diameter can be calculated to $d$ = 6.0 nm, which agrees with the size measured by SEM of 6.0±0.7 nm (**Figure S3b**). The first excitonic transition at 1557 nm renders the NCs perfectly suited for the IR-detection of 1560 nm laser pulses in the third telecommunication window.

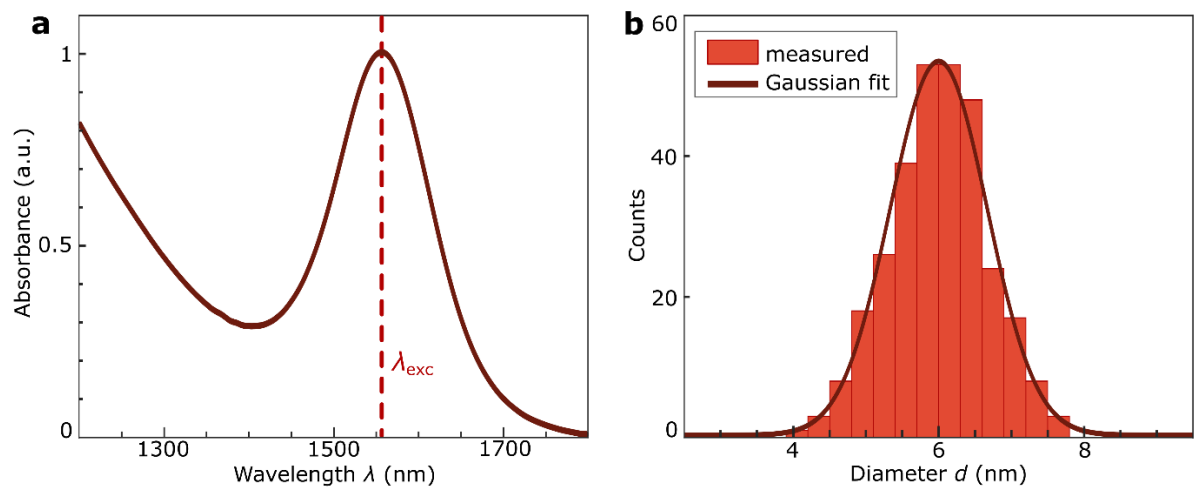

**Figure S3: Characterization of the PbS-NCs.** (**a**) Absorption spectrum of the native PbS-NCs dissolved in tetrachlorethylene. The first excitonic transition appears at 1557 nm ($E_{gap}$ = 796 meV). Applying sizing curves, the NC diameter can be estimated to ~ 6.0 nm.[6] The red bar indicates the position of the laser wavelength for the 2PC excitation of $\lambda_{ex}$ = 1560 nm. (**b**) Size distribution of the NCs derived from scanning electron microscopy (see **Figure S4**). Gaussian fit of the histogram ($n$ > 300) yields a particle size of $d$ = 6.0±0.7 nm, agreeing with the size calculated from (a).

## Supporting Section IV: Characterization of the PbS-NC thin films

We fabricated the photodetectors by spincoating the native PbS-NCs on Si/SiOx devices (200 nm oxide) with prepatterned electrodes in an interdigitated geometry. It was previously shown that Si/SiOx devices are well suited for IR photodetectors.[7] Subsequently, the native OA ligands were exchanged with ethanedithiol (EDT), as detailed in the Methods. The homogeneously coated photodetector channels feature a film thickness of $h$ ~ 50 nm, channel lengths of $L$ = 5.5–15 µm and a total width of $W$ = 1 cm, forming a detector area of ~ 500 × 475 µm² ~ 0.2 mm² (**Figure S4a–c**). Using the channel dimensions, the conductivity of the PbS-NC channels was calculated as $\sigma = (G \times L)/(h \times W)$. The long-chained native OA ligands are replaced by the shorter EDT ligands, resulting a reduced interparticle distance. This in turn enhances the electronic coupling, resulting in a conductive NC ensemble, suitable for photodetection.[8] The success of the ligand exchange was analysed using SEM (**Figure S4d–f**).

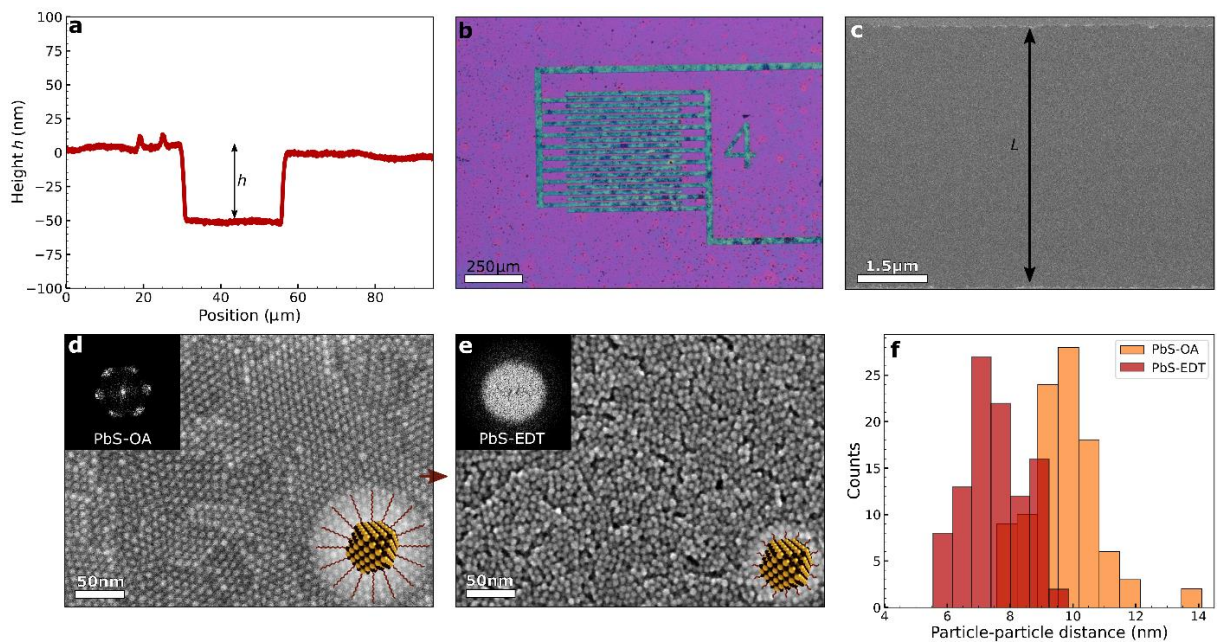

**Figure S4: Characterization of the PbS-NC films.** (**a**) Height profile of the spin-coated PbS-EDT film on the photodetector, measured via profilometry across an induced scratch. The thickness of this sample can be calculated to be $h = 55\pm3$ nm. (**b**) Optical micrograph of a typical PbS-EDT channel prepared by spincoating a $h \sim 50$ nm thick PbS-NC film and subsequent ligand exchange with EDT. The channel length corresponds to $L = 5.5$ µm and the total width of the interdigitated electrodes to $W = 1$ cm, forming a detector area of $\sim 500 \times 475$ µm$^2$ $\sim 0.2$ mm$^2$. (**c**) Scanning electron micrograph of a PbS-EDT channel, highlighting the homogeneous coating of the photodetectors. On the top and bottom edge of the image, the Au electrodes are visible, forming a channel with length $L = 5.5$ µm. (**d**) High-resolution SEM micrograph of native OA-stabilized PbS-NCs. Inset: Corresponding fast Fourier transform, indicating a highly ordered structure with typical sixfold symmetry. A long-range order is observed. (**e**) High-resolution SEM micrograph of the EDT-exchanged PbS-NCs. Inset: Corresponding fast Fourier transform, clearly indicating a successful ligand exchange, based on the missing structural order as a consequence of the volume reduction from long-chained OA to short EDT ligands.[8] (**f**) Distribution of the particle center to center distances for the native PbS-OA (orange) and PbS-EDT (red) NCs (measured via SEM, $n = 100$ each). The inter-particle distance is significantly decreased from $3.73\pm1.1$ nm for OA to $\sim 1$ nm for EDT, agreeing with typical values reported in literature.[9,10]

## Supporting Section V: Native OA-functionalized PbS-NC devices

The native OA-stabilized PbS-NC devices show a highly insulating behavior (**Figure S5a**) and thus no photoresponse to the 1560 nm laser illumination (**Figure S5b**). The exciton recombination takes place within individual photoexcited NCs, as no sufficient electronic coupling between adjacent NCs is present.[8,9,11] However, even if the native PbS-NCs would be photoconductive, a very large response time of $\tau \gg 10$ ns might not be visible with our $\pm5$ ns delay window.[11] Time-resolved photoluminescence and transient absorption measurements on PbS-NCs have revealed drastically reduced exciton lifetime from $\sim 1$ µs to $\sim 6$ ns by exchanging the native OA ligands with EDT.[9,11]

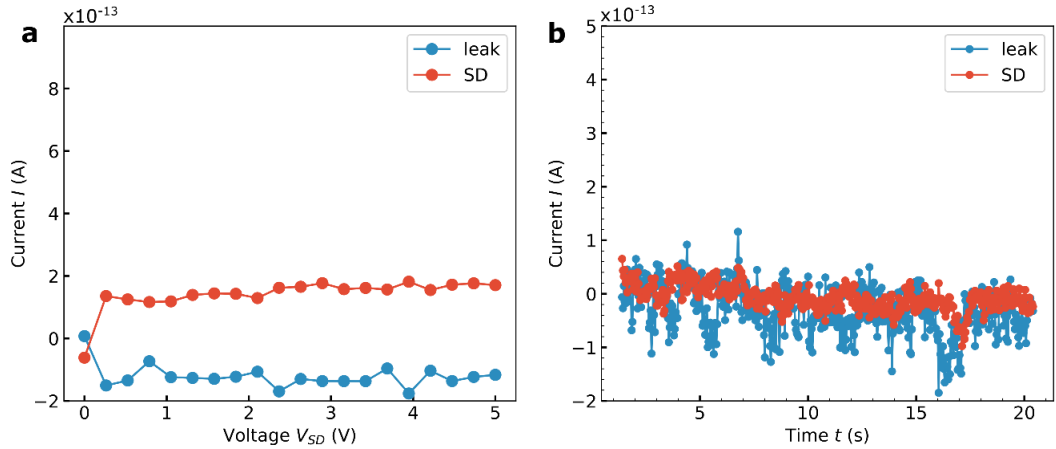

**Figure S5: Insulating character of native PbS-OA channels.** (**a**) Typical *I*-*V*-characteristics of an OA-stabilized PbS-NC channel. Up to $V_{SD} = 5$ V, no significant current flow is observed. (**b**) Current of a PbS-OA channel as a function of time under switching the laser on and off for several times ($P_{average}$ ~ 90 mW). No photoresponse can be identified.

## Supporting Section VI: Probe-induced change in photovoltage

The exemplarily shown 2PC curve of the probe-induced $\Delta V$ in Figure 2b of the main text, the induced change at overlapping pulses corresponds to $\Delta V(\Delta t=0)$ ~ 0.125 mV, whereas the photovoltage at $\Delta t \gg 0$ is $V(\Delta t \gg 0)$ ~ 0.4 V (for $P_{average}$ ~ 90 mW per pulse, $V_{SD} = 1$ V, at room temperature). The values of $V(\Delta t \gg 0)$ are considered to represent the effect of different parameters ($P_{average}$, $V_{SD}$, $V_G$, temperature $T$, channel length $L$) on the photovoltage. The dip at $\Delta t = 0$ corresponds to a change of $\lesssim 1\%$ ($V(\Delta t=0)/V(\Delta t \gg 0)$). For all measurements performed, the change at $\Delta t = 0$ is on the order of ~0.1–1 mV.

The idealized theoretical limit for a full saturation after the first pulse without further effects of the probe pulse corresponds to 50% (c.f. Figure S1). This is clearly not the case for a real system, including our large area detector without focused illumination (c.f. Figure 1d). Experimental 2PC measurements on photodetectors of various materials feature dip modulations ($V(\Delta t=0)/V(\Delta t>>0)$) of a few to several ten percent.[2,12–15] As our PbS-NC photodetectors do not feature an intrinsic electric field, the channels are externally biased, resulting in a voltage offset overlapping with the 2PC curve, which in turn could reduce the percentage of the dip modulation. In addition, it is not yet entirely clear whether the low dip modulation is a result of the PWA sampling via ASOPS. Further, in a previous study, the dip modulation was observed to increase with increasing laser power.[13] Thus, the large area excitation in our setup could also cause the low dip modulation.

Another factor might be the superposition of the response time with the transit and RC time response which occur on much larger time scales and thus reduce the dip modulation of the recombination signal.

However, as the intrinsic response time $\tau$ is not affected by the dip modulation, the determined response times are considered valid.

**Figure S6** displays the 2PC measurements of a PbS-EDT photodetector with the probe pulse off and on, whereas only the latter configuration results in a dip.

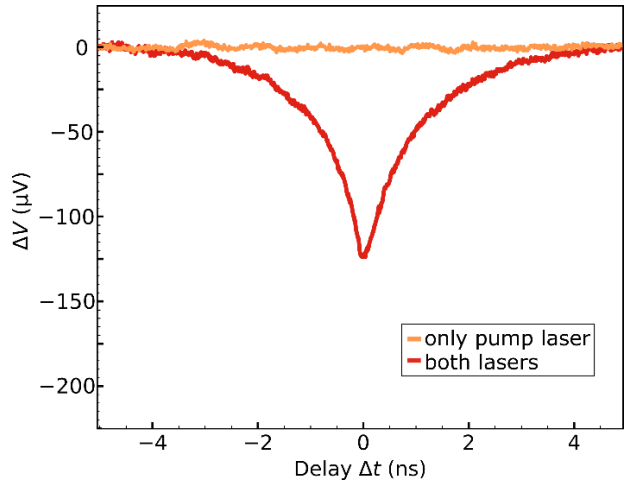

**Figure S6: 2PC measurement with and without probe pulse.** Probe induced signal $\Delta V$ as a function of $\Delta t$ with both pump and probe pulses (red) and only with pump pulse (orange). A characteristic dip is only observed for the pump-probe configuration (red), indicating the probe-induced change of $\Delta V$.

## Supporting Section VII: Discussion of the estimated 3dB bandwidth from $\tau$

Generally, the response time $\tau$ (or time constant) is defined as the time required to decrease in value to $1/e \approx 36.8\ \%$.

For a system where $\tau$ is not limited by the *RC* time constant but the transit time $\tau_{trans}$, the bandwidth can be estimated as $f = \frac{3.5}{2\pi\,\tau_{trans}} \approx \frac{0.55}{\tau_{trans}}$.[16,17] However, this formula is typically used to estimate the intrinsic 3 dB bandwidth (cutoff frequency) from the intrinsic response time, limited by both the transit time $\tau_{trans}$ or the recombination time $\tau_{recomb}$,[12,18–20] to differentiate it from *RC*-limited systems. Thus, for indirect 2-pulse coincidence measurements, this expression should be used to convert $\tau$ into the detector bandwidth to ensure comparability to other materials that have also been characterized using this method.

Additionally, sometimes $\frac{1}{\tau}$ is used to estimate the intrinsic bandwidth.[2,21]

For direct transient photocurrent measurements, often the rise time $t_r$ and fall time $t_f$ are considered, which are defined as the time required for the signal to rise from 10% to 90% or fall from 90% to 10%. The rise time is related to the response time as $t_r = 2.2\tau$. With this, the bandwidth can be estimated from $t_r$ as $f = \frac{2.2}{2\pi t_r} \approx \frac{0.35}{t_r}$.

## Supporting Section VIII: 2PC measurements

**Influence of the source-drain bias $V_{SD}$:**

**Figure S7a** displays the 2PC signal $\Delta V(\Delta t)$ for different source drain biases in the range of $V_{SD} = 0.4–3.5$ V. The response times for different $V_{SD}$ of two devices are given in **Figure S7b** and are roughly constant for all $V_{SD}$. At low $V_{SD}$, the signal-to-noise ratio is too low to achieve proper exponential fitting, resulting in some uncertainty of $\tau$.

The photovoltage $V(\Delta t \gg 0)$ linearly increases with $V_{SD}$, indicating the Ohmic behavior of the PbS-EDT-Au channels. For comparison, the PbS-OA channels show neither a 2PC bleach nor a photovoltage (**Figure S7b**).

**Influence of the gate voltage $V_G$:**

**Figure S7c,d** illustrates the effect of an applied gate voltage on the 2PC response of the PbS-EDT channels. A small increase (~10%) of the response time $\tau$ from 0.9±0.02 ns for $V_G = 0$ V to 1.01±0.05 ns and 1.03±0.01 ns is observed by applying $V_G = -10$ V and +10 V, respectively. The photovoltage $V(\Delta t \gg 0)$ is constant within ~1% for $V_G = \pm 10$ V. This is in good agreement with the transfer characteristics of the PbS-EDT channels under laser excitation (c.f. **Figure S9**), as no current modulation was observed for varying the gate voltage $V_G$. This is consistent with a large free carrier density under photoexcitation.

**Influence of the temperature *T*:**

The 2PC curves of a PbS-EDT photodetector in the temperature range of $T = 10–300$ K under vacuum are plotted in **Figure S7e**. Even at 10 K, a characteristic bleach at around $\Delta t = 0$ is formed. The shape of the bleach appears to be rather similar for all temperatures measured. As shown in **Figure S7f**, $\tau$ is independent of the temperature. The photovoltage $V(\Delta t \gg 0)$ decreases with decreasing *T*, indicating an increase of the resistance *R* with lower *T* due to thermally activated hopping, as expected for PbS-NC thin films.[22,23]

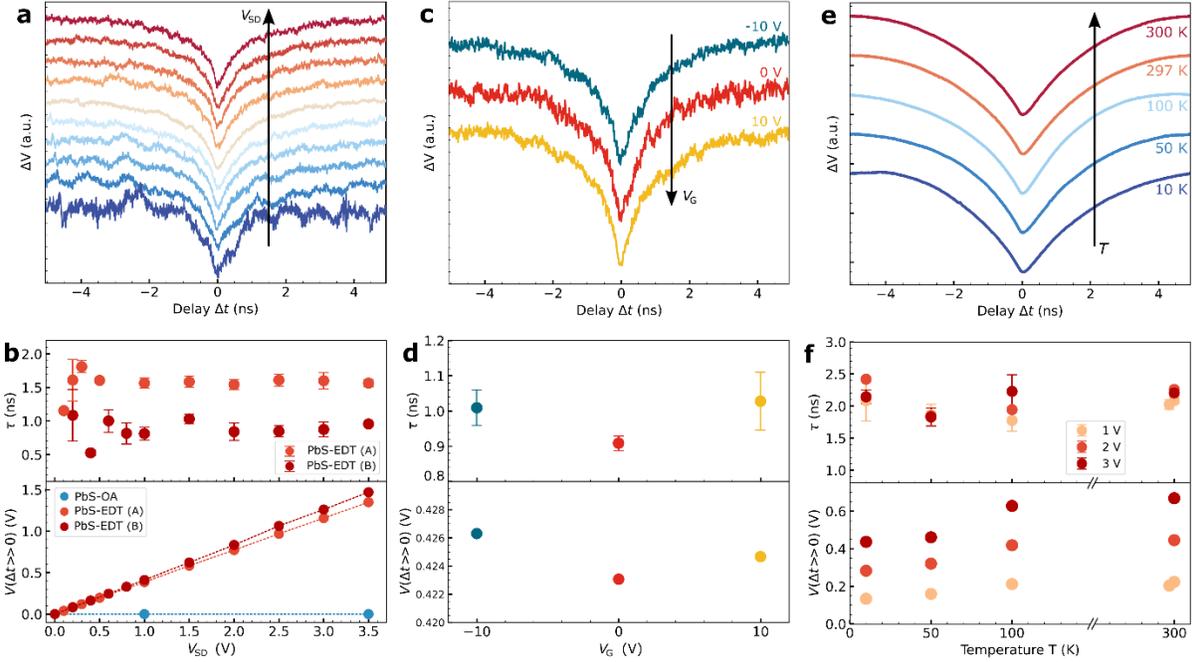

**Figure S7: 2PC measurement results of PbS-NC channels.** (**a**) Normalized 2PC signal $\Delta V(\Delta t)$ for different $V_{SD}$ in the range of 0.4–3.5 V. For 0 V, no characteristic 2PC dip is formed. (**b**) $\tau$ and $V(\Delta t \gg 0)$ as a function of $V_{SD}$. For $\tau$, two different PbS-EDT devices are shown, PbS-EDT (B) corresponding to the data in (a). $\tau$ is independent of $V_{SD}$. $V(\Delta t \gg 0)$ as a function of $V_{SD}$ of two PbS-EDT devices (red). A linear increase of $V(\Delta t \gg 0)$ with $V_{SD}$ is observed. In contrast, no photoresponse is observed for native PbS-OA devices (blue). The lines are guides to the eye. (**c**) 2PC signal for different gate voltages of $V_G$ = -10 V, 0 V, +10 V. $V_{SD}$ = 1 V. (**d**) $\tau$ and $V(\Delta t \gg 0)$ as a function of $V_G$, corresponding to the data in (c). A minor increase (~10%) of the response time from $\tau = 0.9 \pm 0.02$ ns (0 $V_G$) to $\tau = 1.01 \pm 0.05$ ns and $\tau = 1.03 \pm 0.01$ ns is observed by applying -10 $V_G$ and +10 $V_G$, respectively. $V(\Delta t \gg 0)$ stays constant within ~1% for ±10 $V_G$. (**e**) 2PC curves of a PbS-EDT channel at different temperatures of $T$ = 10–300 K. (**f**) $\tau$ and $V(\Delta t \gg 0)$ as a function of $T$ for different $V_{SD}$, corresponding to the data in (e). $\tau$ is insensitive to $T$. $V(\Delta t \gg 0)$ decreases with decreasing $T$. Measurements in (a–d) were performed under $N_2$ at room temperature, measurements in (e,f) under vacuum. $V_{SD}$ = 1 V, $V_G$ = 0 V and $P_{average}$ ~ 90 mW, unless otherwise specified. The error bars indicate the fitting uncertainty. The curves in (a,c,e) are normalized and offset for clarity, $V(\Delta t \gg 0)$ values are extracted before normalization.

**Influence of the channel length $L$:**

We characterized several samples with different channel lengths $L$ to investigate its effect on the response time $\tau$ (**Figure S8**). We observe that the response time $\tau$ is independent on the channel length $L$, whereas the photovoltage $V(\Delta t \gg 0)$ increases with decreasing $L$, as expected due to the decreased resistance.

The above-mentioned response times vary slightly from sample to sample, but the observed trends are the same.

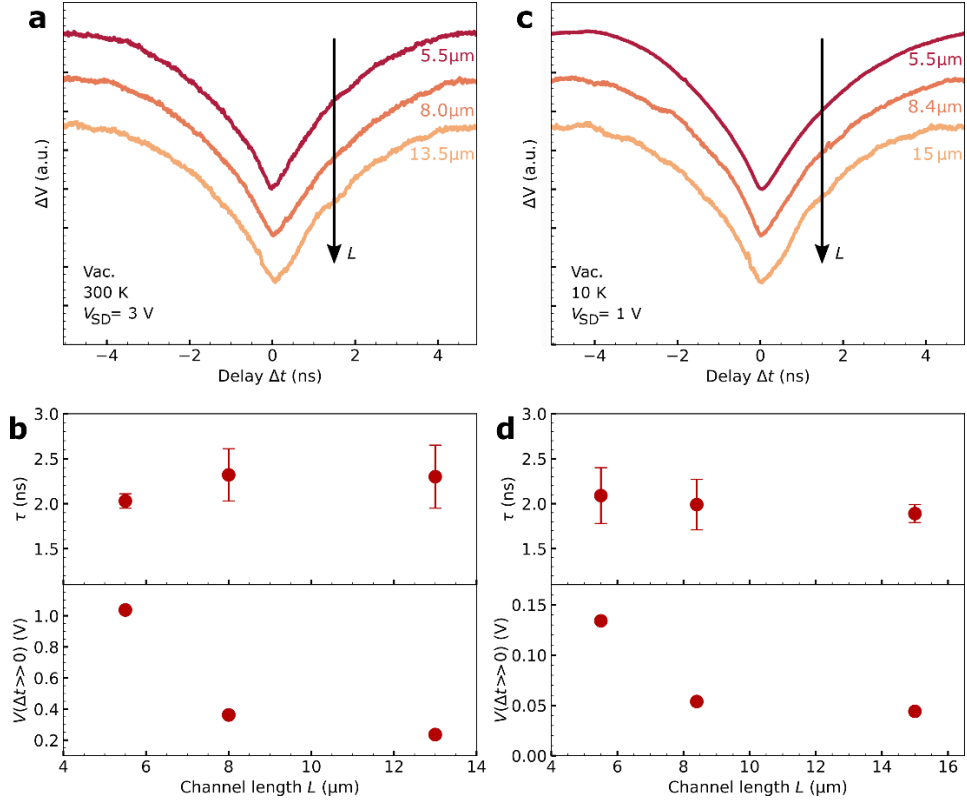

**Figure S8: 2PC measurements of PbS-EDT photodetectors with different channel lengths.** (**a,c**) 2PC signal $\Delta V$ as a function of $\Delta t$ for different channel lengths of $L$ = 5.5–15 µm for (**a**) $V_{SD}$ = 3 V, $T$ = 300 K and (**c**) $V_{SD}$ = 1 V and $T$ = 10 K, respectively. The curves are normalized and offset for clarity. (**b,d**) $\tau$ and $V(\Delta t \gg 0)$ as a function of $L$, corresponding to the data in (a) and (c), respectively. No clear trend for $\tau$ is apparent, whereas $V(\Delta t \gg 0)$ decreases with increasing $L$. The error bars indicate the fitting uncertainty. Measurements were performed under vacuum at the indicated $T$ and $V_{SD}$, $V_G$ = 0 V, $P_{average}$ = ~90 mW. $V(\Delta t \gg 0)$ values are extracted before normalization.

## Supporting Section IX: Gating of the PbS-NC channels

As the PbS-EDT photodetectors are prepared on a Si/SiOx device, field-effect transistor (FET) measurements can be performed. **Figure S9** displays the transfer curves of the PbS-EDT channels in the dark (**Figure S9a**) and under laser illumination (**Figure S9b**), respectively.

In the dark, a semiconducting p-type behavior is observed. To estimate the field-effect mobility of the holes $\mu(h^+)$ of the PbS-NC channels, the gradual channel equation is used, given in Equation (S2):[10]

$$\mu = \frac{\partial I_{SD}}{\partial V_G} \frac{L}{W} \frac{t_{ox}}{\varepsilon_0 \varepsilon_r V_{SD}} \quad, \tag{S2}$$

$\frac{\partial I_{SD}}{\partial V_G}$ corresponds to the derivation of the transconductance curve ($I_{SD}$ as the detected source-drain current and $V_G$ as the applied gate voltage). $L$ and $W$ are the channel length and width. $t_{ox}$ and $\varepsilon_0 \varepsilon_r$ are the thickness (200 nm) and the permittivity of the dielectric SiOx layer, respectively. $V_{SD}$ corresponds to the applied source-drain voltage.

The hole mobility is estimated as $\mu(h^+) \sim 4 \times 10^{-5}$ cm$^2$/Vs, agreeing with reported values in literature.[22,24] Under laser illumination, no field effect is present, indicating a high carrier density.

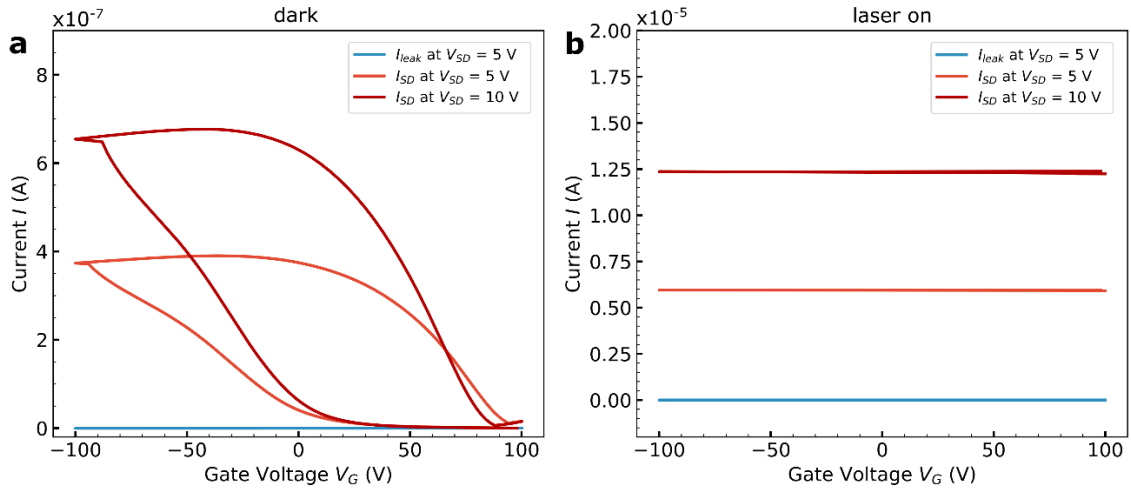

**Figure S9: FET transfer characteristics of the PbS-NC channels.** (**a**) Typical transfer curves of a PbS-EDT channel in the dark for different source-drain biases of $V_{SD}$ = 5 V and 10 V, respectively. P-type behavior is observed. The hole mobility can be calculated to $\mu(h^+) \sim 4 \times 10^{-5}$ cm$^2$/Vs. The leak current is negligible. (**b**) Transfer curve of the same channel under pulsed laser (quasi-CW) illumination with $P_{average}$ ~ 90 mW. No modulation of $I_{SD}$ with modulating $V_G$ is observed. The leak current is negligible.

## Supporting Section X:
## Influence of air exposure on the intrinsic response time τ

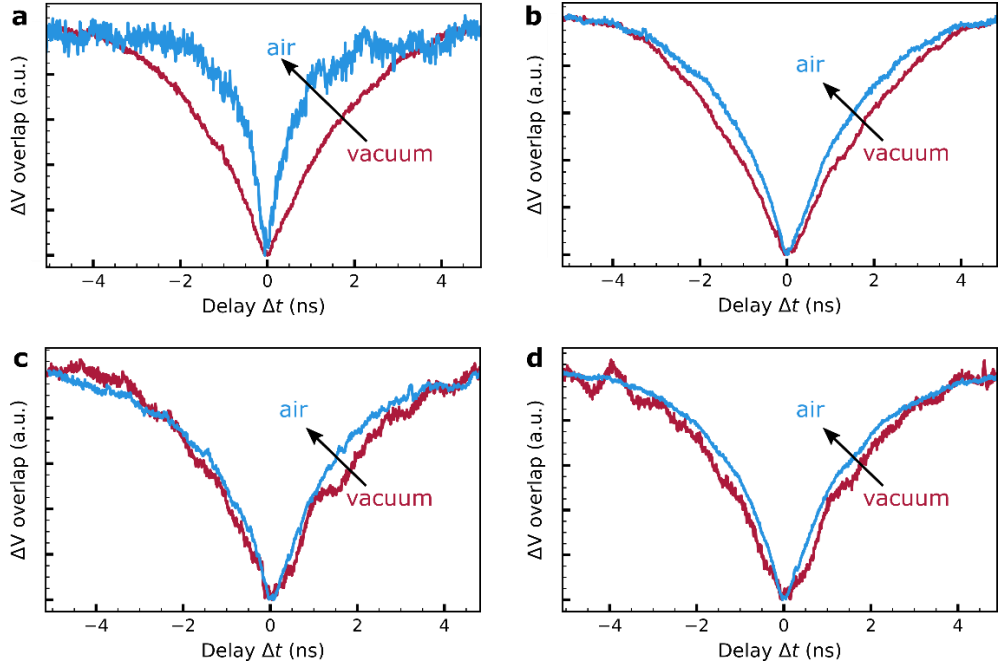

**Figure S10: Influence of air exposure on the intrinsic response time τ.** (**a–d**) 2PC curves of four different PbS-EDT photodetectors under vacuum (red) and after air exposure, measured under $N_2$ (blue). For all channels, a sharper 2PC dip and thus faster τ is observed after air exposure (τ improved Ø by factor of 1.88) and $V(\Delta t \gg 0)$ significantly increased (factor of 2.01). Details: From a–d, the intrinsic response time τ improved by a factor of 3, 1.51, 1.46 and 1.51, whereas the photovoltage $V(\Delta t \gg 0)$ increased by a factor of 1.4, 3.6, 1.4 and 2 after exposure to air. The channel lengths correspond to $L = 5.5$ µm, 13 µm, 13 µm and 8 µm, for a, b, c and d, respectively. The curves are normalized to overlap at $\Delta t \gg 0$ and $\Delta t = 0$ for direct comparability, $V(\Delta t \gg 0)$ values are extracted before normalization.

## Supporting Section XI: Estimation of the carrier concentration

The carrier concentration within the PbS-EDT films can be roughly estimated using the Equation (S3):

$$n = \frac{\sigma}{e \times \mu}, \quad (S3)$$

with the elementary charge $e = 1.602 \times 10^{-19}$ C, the conductivity $\sigma_{\text{dark}} = 6 \times 10^{-6}$ S/m, $\sigma_{\text{photo}} = 7 \times 10^{-3}$ S/m, and the hole mobility $\mu(\text{h}^+) = 4.4 \times 10^{-5}$ cm$^2$/Vs. The latter corresponds to the mobility determined for PbS-EDT channels under dark conditions, as under pulsed laser excitation, a field-effect was absent. However, it can be assumed that the laser excitation does not significantly change the carrier mobility and just increases the carrier concentration,[9] responsible for the higher conductivity. Thus, the carrier concentration in dark and under photoexcitation can be estimated as $n_{\text{dark}} \sim 10^{16}$ cm$^{-3}$ and $n_{\text{photo}} \sim 10^{19}$ cm$^{-3}$, respectively.